\documentstyle[draft]{npbproc}

\newcommand{\be}{\begin{equation}}
\newcommand{\ee}{\end{equation}}

\begin{document}

\title{Instantons in the quantum framework of 2D  Gravity \thanks{Supported by
CICYT,Spain}} \author{J.Luis Miramontes\thanks{Address after october 1, CERN,Th
Division.}} \address{Department of Physics: Joseph Henry Laboratories,
Princeton University, Princeton, NJ 08544}
\author{J.Sanchez Guillen\thanks{e-mail:joaquin@euscvx.cern.ch}}
\address{Department of Particle Physics, University of Santiago, E-15706
Santiago, Spain}
\runtitle{Instantons}
\runauthor{Miramontes, Sanchez Guillen}
\begin{abstract}
We analyze the non--perturbative features of 2D quantum gravity
defined by stochastic regularization of the
unstable matrix model showing, first, that the WKB approximation of
the well-defined quantum Fokker-Planck hamiltonian corresponds to the
semiclassical eigenvalue density of the former. The double scaled potential
exhibits an instanton--like behaviour, which is universal and scales, but whose
interpretation in terms of pure gravity is still open. \end{abstract}
\maketitle
1) INTRODUCTION

In the last year considerable effort has been devoted to solve the non--
perturbative ambiguities in the possible definition of 2D quantum gravity
trough matrix model techniques. While the solution\cite {BPM} for the $k=3$($k$
 odd in
general) multicritical model is better understood \cite{Lykken}, the problem
for
 the $
k$ even models is much more complicated and it is still open whether the
non--perturbative features of one--matrix models are related to pure gravity at
all \cite{Ambjorn}. The difficulty is related to the instability of the
matrix model with potentials unbounded from below, an old problem of euclidean
gravity for which the two classical cures have been tried also here.
The contour deformation \cite{David}  leads
to a complex solution, presumably related
to the complex Borel sum of the perturbative expansion, whose imaginary part is
proportional to the non--perturbative ambiguity obtained from the linearized
 Painlev\'e
equation and has been interpreted as an instanton effect \cite{Ginsparg}. The
 second
method is stochastic regularization, where one converts the ill defined
$D=0$ problem into a well defined $D=1-like$ one, and so it is considered
by many to be the only realistic hope for a consistent real solution, albeit
suspect as a possible source of additional arbitrariness.So far, only numerical
computations have been attempted \cite{AmbG}, some of them  related to the
supersymmetry breaking\cite{Geli} , which  is a very interesting aspect of the
 original
proposal \cite{Parisi} directly related to the $D=1$ properties which will not
 be
discussed here.

Most of the numerical results rely on the decoupled N--fermion formulation,
 which turns
out to be a property of potentials of degree $<4 $. We prove here also that
 a mean field  may be introduced in the WKB approximation to decouple the
 system,
so that the semiclassical limit of the Matrix Model is reproduced. This way,
the
 $D=1$
system becomes an ideal Fermi gas whose spectral density is expressed
in terms
of the Fokker--Planck potential.

We are then able  to investigate analytically the leading non--perturbative
behaviour of the stochastic stabilized quantum gravity, also in the WKB
approximation. We find that the double scaled (universal) stabilized Fokker--
Planck potential  exhibits in the $k-even$ case instanton--like effects,
related
to the non--Borel summability, and corresponding to the lifetime of a
metastable
 state
which is universal and characteristic of the non--perturbative contributions.
On
the other hand there are no such effects in the $ k=3 $ multicritical
 potentials,
as one should expect.

On the light of this  result we comment the previous numerical attempts
and discuss  the leading effect of this tunnelling ($bounce$) contribution of
 the  $D=1
- like$ spectrum in the free energy of the $D=0$ model, which can have
different
interpretations.

\vspace{.5 cm}

\noindent
2) WKB APPROXIMATION AND THE SEMICLASSICAL MATRIX MODEL

Let us consider the $D=0$ hermitean one--matrix model defined by the partition
function
\begin{equation}
Z=\int D\phi e^{-\beta V(\phi)}
\label{eq:partition}
\end{equation}
where $\phi$ is an $N\times N$ hermitean matrix, and $V(\phi)$ is a generic
potential of degree $L$:
\begin{equation}
V(\phi)  = \sum_{n=2}^{L} g_n Tr (\phi^n)
\label{eq:potential}
\end{equation}
This model can be understood in terms of the ground state
of a $D=1$ quantum mechanical system, whose hamiltonian is the (positive
semi--definite) Fokker--Planck hamiltonian \cite{Amb,Noso}
\begin{eqnarray}
H_{FP}&=& Tr(P^2) + W_{FP} \quad ;
\qquad P_{ij}\;=\;-i{\partial\over\partial \phi_{ji} } \nonumber\\
W_{FP}&=& {\beta^2\over4} {\partial V\over\partial \phi_{ij}} {\partial V\over
\partial \phi_{ji}} - {\beta\over2} {\partial^2 V\over \partial \phi_{ij}
\partial \phi_{ji} }
\label{eq:HFP}
\end{eqnarray}
This means that the expectation value of an operator $Q(\phi)$ in the matrix
model can be defined as the vacuum--expectation--value (VEV) of the quantum
operator $Q(\hat \phi)$
\begin{eqnarray}
&\langle Q \rangle \equiv \langle 0 \vert Q(\hat \phi) \vert 0 \rangle
= \int D\phi \Psi_0^2(\phi) Q(\phi)& \nonumber\\
&H_{FP} \Psi_0 (\phi) = E_0 \Psi_0(\phi) &
\label{eq:QMdef}
\end{eqnarray}
where $E_0$ and $\Psi_0(\phi)$ are the energy and the wave function of the
ground state. $E_0 = 0$ corresponds to the case when the potential
is bounded from below, the matrix model is well defined, and $\Psi_0(\phi) =
exp(-\beta V(\phi)/2)$; while eq.\ref{eq:QMdef}, with $E_0 >0$, defines
$\langle Q \rangle$ in terms of the true ground state wave function.

Following  well known techniques \cite {BIPZ}, it is natural to make a
change of variables to the eigenvalues $\lambda_i$ of the matrix $\phi$, by
introducing the effective ground state wave function
\begin{equation}
\Psi_0^{eff} (\{\lambda_i\}) = \prod_{i<j} \left( \lambda_i - \lambda_j \right)
\Psi_0 (\phi)
\label{eq:antisymm}
\end{equation}
$\Psi_0^{eff}$ is totally antisymmetric, and it describes a gas of $N$ Fermi
particles.
In general, the Fokker--Planck potential, $W_{FP}$, can be splitted into its
diagonal ($D$) and non--diagonal ($ND$) parts, which after diagonalization
of the matrix read %
\begin{eqnarray}
W_{FP}  &=&  W_{FP}^{(D)} + W_{FP}^{(ND)} \\
W_{FP}^{(D)}&=&{\beta^2\over4} \sum_{i=1}^N
\left((V'(\lambda_i))^2 - \right. \nonumber\\
& & \left. - 4 X \left( g_2 + \sum_{n=3}^L
n g_n \lambda_i^{n-2}\right)\right) \nonumber\\
W_{FP}^{(ND)} &=& -{\beta\over2} \sum_{i,j=1}^N
\left(\sum_{n=4}^L n g_n \sum_{s=0}^{n-4}
\lambda_i^{s+1}\lambda_j^{n-3-s} \right) \nonumber
\label{eq:potencialFP}
\end{eqnarray}
Where $X=N/\beta=e^{\gamma_0}$ is related to the (bare) 2D cosmological
constant in the usual way. Obviously the $ND$ piece does not vanish in general
if $L\ge 4$, and the Fermi gas is not decoupled. Nevertheless, in the
semiclassical WKB approximation
($\beta\approx \hbar^{-1}\rightarrow\infty$), a mean field
approximation (\`a la Hartree--Fock) may be performed to decouple the
system. We show below that
\begin{eqnarray}
Tr(\phi^k) Tr(\phi^p) &\approx&
N \left( \omega_k Tr(\phi^p) + \omega_p Tr(\phi^k) - \right. \nonumber\\
& & \left. - N \omega_k \omega_p \right) + \cdots
\label{eq:Hartree}
\end{eqnarray}
where the normalization is fixed by
\begin{eqnarray}
&\langle Tr(\phi^k) Tr(\phi^p) \rangle_c \approx N^2 \left(\omega_k \omega_p +
O(1/N)\right) \; ; & \nonumber\\
&\omega_k = {1\over N} <Tr(\phi^k)>&
\label{eq:normaliza}
\end{eqnarray}
is consistent with the semiclassical limit of the matrix model.

Under eq.\ref{eq:Hartree}, the quantum mechanical system becomes an ideal Fermi
gas of N particles, with the hamiltonian
\begin{eqnarray}
H_{FP}  &\approx& \sum_{i=1}^N h^{FP} (\lambda_i) \nonumber\\
h^{FP}(\lambda)\;&=&\; -{\partial^2\over\partial\lambda^2} +
{\beta^2\over4} U_{FP} (\lambda) \\
U_{FP} (\lambda) &=& - 4X\left( \sum_{i=1}^{L-2}
\left( \sum_{j=i+2}^L jg_j \omega_{j-i-2}\right) \lambda^i \right. \nonumber\\
\lefteqn{\left. + g_2 - {1 \over 2}
\sum_{i=4}^L ig_i \sum_{j=2}^{i-2} \omega_{j-1} \omega_{i-j-1} \right) +
\left( V'(\lambda)\right)^2} \nonumber
\label{eq:fermion}
\end{eqnarray}
Notice that the  potential $U_{FP}$ is bounded from below, and the
one--fermion
hamiltonian has a well--defined discrete spectrum, $h^{FP} \phi_n(\lambda)
= e_n \phi_n(\lambda)$. All the relevant information
about the $D=1$ quantum mechanical model is contained in the density,
$\rho(\lambda,e)= \langle \lambda \vert \delta \left(h^{FP} - e \right)\vert
 \lambda
\rangle$, and its normalization fixes the Fermi energy, $e_F$,
\be
N = \int d\lambda\ \int_{-\infty}^{e_F} de\ \rho(\lambda,e)
\label{eq:fermienergy}
\ee
The energy integral of $\rho(\lambda,e)$
provides the quantum mechanical version of the semiclassical density of
eigenvalues in the matrix model
$$
u (\lambda) = {1\over\beta} \int_{-\infty}^{e_F} de\ \rho(\lambda,e)
\; ; \; X= {N\over\beta}=\int d\lambda\ u (\lambda)
$$
which, in the WKB approximation, reads
\begin{eqnarray}
u^{WKB} (\lambda) = {1\over2\pi} \sqrt{e_F - U_{FP}(\lambda)}\ \theta(
e_F - U_{FP}(\lambda)) \nonumber\\
{1\over\beta}\ \langle Tr\ Q(\phi) \rangle = \int
d\lambda\ Q(\lambda)\ u^{WKB} (\lambda)
\label{eq:densityWKB}
\end{eqnarray}

Let us go back to the $D=0$ matrix model. We shall use the Schwinger--Dyson
loop equations \cite{LDavid} to get a detailed expression for the
semiclassical density of eigenvalues, and compare with the WKB approximation of
the quantum mechanical system, eq.\ref{eq:densityWKB}.
In the semiclassical (planar) limit, the generating function of monomial
expectation values,
\be
F(p) = {1\over\beta} \langle Tr {1\over p-\phi}\rangle_c
\label{eq:generator}
\ee
satisfies the Schwinger--Dyson equation
\begin{eqnarray}
& &F(p)^2 - V'(p) F(p) + \nonumber\\
& &+ X \sum_{i=0}^{L-2}
\left( \sum_{j=i+2}^L j g_j \omega_{j-i-2} \right) p^i = 0
\label{eq:loop}
\end{eqnarray}
where $\omega_k$ are the ``constants of integration'' of this loop equation,
and have already been defined in eq.\ref{eq:normaliza}. Therefore,
eq.\ref{eq:loop} is
solved as
\begin{eqnarray}
F(p) &\equiv& \int d\lambda {u^{SC}(\lambda)
\over p-\lambda} \label{eq:f(p)} \\
&=&{1\over 2}\left( V'(p) - \sqrt{\Delta(p)}\right) \nonumber\\
\Delta(p) &=&
\left(V'(p)\right)^2 - 4X \sum_{i=0}^{L-2} \left( \sum_{j=i+2}^L j g_j
\omega_{j-i-2} \right) p^i \nonumber
\end{eqnarray}
The ``constants of integration'', $\omega_k$, are
fixed by the condition that the imaginary part of $F(p)$ defines a proper
semiclassical density of eigenvalues, $u^{SC} (\lambda)$. Obviously,
$\Delta(p)$ is a polynomial in $p$ and all the
branch cuts of eq.14 will be squared root branch cuts. Therefore,
$\Delta(p)$ has to satisfy the following constraints \cite {BIPZ,David} :
(i) $\Delta(p)$ must have only real zeros in the complex $p$--plane,
and (ii) $\Delta(p)$ cannot have three consequtive odd degree zeroes.
Under these conditions, the semiclassical density of
eigenvalues is given by
\begin{eqnarray}
u^{SC} (\lambda)&=& {1\over\pi} Im\ F(\lambda) \nonumber\\
&=& {1\over2\pi} \sqrt{-\Delta(\lambda)}\ \theta(-\Delta(\lambda))
\label{eq:densitySC}
\end{eqnarray}
The square root branch cuts of $F(\lambda)$, i.e., the intervals between odd
degree
(real) zeros of $\Delta(\lambda)$, are the ``bands'' on which $u^{SC}(\lambda)$
has support.

Therefore, under the above mentioned restrictions, the comparison between
eqs.(14), (\ref{eq:densitySC})
and (9), (\ref{eq:densityWKB})
shows that the WKB limit of
the $D=1$ Fokker--Planck hamiltonian, with the mean field approximation of
eq.\ref{eq:Hartree}, reproduces the semiclassical limit of the matrix model.
Besides, the Fermi energy can be identified in terms of the $\omega_k$
\begin{eqnarray}
{e_F^L \over 4  X} &=& g_2 + 3g_3 \omega_1 + \\
& &+\sum_{i=4}^L \left( \omega_{i-2} +
{1\over2} \sum_{j=2}^{i-2} \left( \omega_{j-1} \omega_{i-j-1}\right) \right) i
g_i \nonumber
\label{eq:Fermiomega}
\end{eqnarray}
This result particularizes for the $D=0$ hermitean one--matrix model the
general arguments about the stabilization of bottomless
euclidean field theories \cite{GH}. It also agrees with, and generalizes,
 previous
results obtained for the simplest potentials with $L=3$
 \cite{Parisi,Amb,Noso} and
$L=4$ \cite{Geli}, showing that the critical behaviour of the Fokker--Planck
hamiltonian is precisely that of the matrix model.

Therefore, it is possible to describe
the hermitean matrix model like an ideal Fermi gas of N particles, whose
one--fermion potential is fixed by the semiclassical density of eigenvalues
\be
U_{FP}(\lambda) = e_F - \lbrack 2\pi u^{SC} (\lambda)\rbrack^2
\label{eq:potdens}
\ee
This relationship formally holds only when the above mentioned constraints on
$\Delta=U_{FP}-e_F$ are satisfied, and $u^{SC}$ is well defined.
Nevertheless, the quantum mechanical system, and $u^{WKB}$, is defined even
when this is not the case. Such quantum mechanical configurations arise when
the {\it naive} ground state wave function, $\Psi_0= exp(-\beta V/2)$, is not
normalizable in the WKB approximation.
Therefore, they are related to the true ground state with $E_0>0$.

\vspace{.5 cm}
\noindent
3) CONTRIBUTION OF METASTABLE\\
STATES AND INSTANTONS

Once we have shown that the semiclassical density of eigenvalues is just the
WKB
density of the Fokker--Planck hamiltonian, we can use the matrix model results
to get information about the Fokker--Planck potential $U_{FP}$. In particular,
we want to find the features of $U_{FP}$ that are directly related to the
critical behaviour and, therefore, which will survive after the double
scaling limit.

The main problem of the matrix model for pure gravity is that the
susceptibility
which should be determined by the Painlev\'e equation is only given as an
asymptotic series, which is non-Borel summable and so the difference between
any Borel sum (corresponding to different boundary conditions) and the
hypothetical exact solution  is proportional to $T^{-1\over
 8}e^{{-4\sqrt{6}\over
5}T^{5\over 4}}$ . The exponential factor is an instanton-like effect as
 expected
from the large order behaviour of the series, and it has been analyzed in the
semiclassical limit of the matrix model and interpreted as the imaginary part
of
 the
complex Borel sum \cite{Ginsparg}, related to the obstruction of real solutions
\cite{David}, with the conclusion that this matrix model does not define the
sum
 over
topologies\cite{Zinn}.

The first attempts to understand this problem in the stabilizing Fokker--Planck
\cite{AmbG} was the computation of the $D=0$ free energy, which was obtained
differentiating with respect to a source term  added to the Fokker--Planck
 potential
and nothing related to the instability was observed in the numerical results.
 This
motivates further the present attempt to investigate these non--perturbative
 aspects
analytically.

Let us start by computing $\langle F(p) \rangle_c$,
eqs.\ref{eq:generator}, 14, using orthogonal
polynomials \cite{GM} (we restrict
ourselves to the case of even potentials):
\begin{eqnarray}
\langle F(p) \rangle_c &=& \\
\lefteqn{={1\over\beta} \langle Tr {1\over p - \phi}
\rangle = \sum_{n=0}^{N-1} \frac {1}{\beta h_n^2} \int d\lambda e^{-\beta
V} (p-\lambda)^{-1} P_n^2} \nonumber\\
\lefteqn{\equiv {1\over\beta} \sum_{n=0}^{N-1} \langle n \vert{1\over
 p-\hat\phi} \vert
n \rangle= {1\over p\beta} \sum_{n=0}^{N-1} \sum_{j=0}^{\infty}{1\over p^j}
 \langle n
\vert \hat\phi^j \vert n \rangle }\nonumber\\
\lefteqn{= {1\over p\beta} \sum_{n=0}^{N-1} \sum_{j=0}^\infty {{2j}\choose j}
 \left(
R(n) \over p^2\right)^j} \nonumber\\
\lefteqn{= {1\over\beta} \sum_{n=0}^{N-1} {1\over \sqrt{p^2 - 4 R(n)}}
 \rightarrow
\int_0^X dx {1\over \sqrt{p^2-4R(x)}}} \nonumber
\label{eq:polynomials}
\end{eqnarray}
Taking the imaginary part, we get the semiclassical density of
eigenvalues \cite {BIPZ}
$$
u_{SC} (p) =\frac{Im \langle F(p) \rangle_c} {\pi} = {1\over \pi}\int_0^X \! dx
\frac{\theta( 4R(x) - p^2)} {\sqrt{4R(X)-p^2}}
\label{eq:densitypol}
$$
Next, we make the change of variables, $x=W(R(x))$, with $W(R)$
defined in the usual way in terms of the potential
$$
W(R)=\oint {dz\over 2\pi i} V'(z+{R\over z})
$$
an the final expression reads
\begin{eqnarray}
u_{SC} (p) &=& {1\over\pi} \int_{p^2/4}^{R(X)} dR {W'(R) \over \sqrt{4R-p^2}}
\theta( 4R(X) - p^2) \nonumber\\
\lefteqn{ \equiv {1\over2\pi} \sqrt{e_F -U_{FP}(p)} \theta( e_F -U_{FP}(p)) }
\label{eq:densityPOLY}
\end{eqnarray}
As a first result, we see that the semiclassical range for the eigenvalues
is given by $p^2\le 4R(X)$.

Now we can obtain the critical behaviour of $u_{SC}(p)$ in the double scaling
limit. For the $k$mo\-del it corresponds to
\footnote{Notice that $X=W(R)$ and, with these general defintions,
$$
f(T) \simeq {1\over R_c} \left(T\over\gamma\right)^{1\over k} + \ldots
$$}
\be
\begin{array}{rcl}
W(R) &=& X_c \left( 1- \gamma (R_c -R)^k \right) \\
X &=& X_c \left( 1- \beta^{-{2k\over2k+1}} T \right) \\
R &=& R_c \left( 1 - \beta^{-{2\over2k+1}} f(T)\right)
\end{array}
\label{eq:doublesc}
\ee
Therefore,
\be
u_{SC} (p) = {2k\gamma\over \pi2^{2k}} X_c \int_0^{^{\sqrt{4R(X) - p^2}}}
\!\!\!\!\!\!\!\!\!\!\!\!\!\!\!\!\!\!\!\!\!\!\!\!\! dy
\left( 4R_c -p^2 - y^2\right)^ {k-1}
\label{eq:ufinal}
\ee
Notice that, from eqs.\ref{eq:densityPOLY}, \ref{eq:ufinal},
$U_{FP}(\lambda)$ has a root
of order $2k-1$ at $\lambda^2= 4R_c$, in the critical point, $X=X_c$ and
 $R=R_c$.

Let us specialize eq.\ref{eq:ufinal} for $k=2$, pure gravity. Then,
\begin{eqnarray}
u_{SC} (p) &=& {\gamma\over4\pi} X_c \left[ (4R_c - p^2) \sqrt{4R(X) -
 p^2}\right.
-\nonumber\\
& &\left.-{1\over3} (4R(X)-p^2)^{3\over2} \right]
\label{eq:ufinalpg}
\end{eqnarray}
This means that the critical behaviour of the Fokker--Planck potential is
\begin{eqnarray}
U_{FP}(\lambda) - e_F &=& \\
\lefteqn{= \left(\gamma X_c \over 3\right)^2 \left(
\lambda^2-4R(X) \right) \left( 6R_c -2R(X) - \lambda^2\right)^2} \nonumber\\
\lefteqn{ \approx \! \left(\gamma X_c \over 3\right)^2 \!\!\!
 \left((\lambda^2-4R_c)^3 -
\frac {12 \beta^{-{4\over5}}}{\gamma} (\lambda^2-4R_c)T + \cdots \right) }
 \nonumber
\label{eq:potentialcrit}
\end{eqnarray}
Now, the leading behaviour of the potential around $\lambda_0=\pm\sqrt{4R_c}$
when $\beta\rightarrow\infty$ is

\be
U_{FP}(\lambda) \approx \pm 16 \beta^{-{6\over5}} \gamma X_c^2
\left( \frac{4\gamma R_c z^3}{9} \! - \frac{Tz}{3}\right)
\label{eq:potfinal}
\ee
where $z=\beta^{2\over5}(\lambda-\lambda_0)$. Therefore, the double scaling
limit of  $U_{FP}$ has two (symmetric) degenerated secondary
minima at $z_m^2 = \pm \sqrt{T/(4\gamma R_c)}$, if $T>0$, and one absolute
minimum at $\lambda=0$.

The perturbative expansion around $\lambda=0$, the absolute minimum of the
potential, corresponds to the semiclassical WKB expansion, which, as
it is well known, reproduces the $1/N$ expansion of the matrix model\cite{GH}.
In this expansion, the subdominant terms in eq.24 are crucial because
the  normalization condition of having $N\rightarrow\infty$ bound states
depends
also on $\beta$ (in fact, these terms ensure that the potential has bound
states!).
Nevertheless, there is also a non--vanishing contribution related to the
secondary minima which implies the existence of a metastable state,
corresponding precisely to the Fermi level, which
decays due to barrier penetration, and whose non--perturbative
lifetime remains finite in the double scaling limit.
Such lifetime is propotional
to the inverse of the imaginary part of the metastable state energy, as
we show below.

The dominant barrier penetration contribution can be computed in the
semiclassical approximation \cite{Zinn}.We consider the hamiltonian

$ h^{FP}(\lambda) = -
{\partial^2\over\partial\lambda^2} + {\beta^2\over4} U_{FP} (\lambda)
$. The corresponding {\it euclidean} partition function can be written as the
$L\rightarrow\infty$ limit of the functional integral
\be
tr \left( e^{-LH}\right) = \int_{q(-L/2)=q(+L/2)} \lbrack d q(t) \rbrack
e^{- S[q(t)]}
\label{traza}
\ee
where $S[q(t)]$ is the euclidean action
$$
S[q(t)] = \int_{-L/2}^{+L/2} dt \left({1\over4} \dot q^2(t) + {\beta^2\over 4}
U(q(t)) \right) \label {accioneucl}
$$
and $tr(e^{-LH})$ is given by $\sum\ exp{-LE_n}$, where the sum extends over
the
whole spectrum. Therefore, in the $L\rightarrow\infty$ limit,
\begin{eqnarray}
Im\ tr(e^{-LH}) &\propto& Im\ e^{-L(E_{m}^{r}\ +\ iE_{m}^{i})} \nonumber\\
&\propto& L e^{-LE_{m}^{r}} E_{m}^{i}
\end{eqnarray}
where he have taken into account the smallness of the, non--perturbative,
imaginary part of the energy $ E_m^i$ of the lowest eigenvalue that becomes
 complex.

We assume that the potential has a secondary minimum at $q=q_m$, which
obviously requires that the
equation $U(q)=U(z_m)$ has, at least, another solution; we shall call $q_0$
the one that is closest to the minimum, assuming $q_0<q_m$. Therefore, the
would--be eigenvalue corresponding to the
secondary minimum becomes complex, providing  the dominant
contribution to $Im\ tr(e^{-LH})$. Its imaginary part is given by an
instanton, which is a solution of the euclidean equations
of motion that starts from the secondary minimum at euclidean {\it time}
$L=-\infty$, is reflected in $q_0$, and comes back to the minimum at {\it time}
$L=+\infty$.

The euclidean equation of motion is
\be
{1\over2} \ddot q_c = {\beta^2\over4} U'(q_c)
\label {motion}
\ee
which has a conserved quantity
\be
\dot q^{2}_{c} -\beta^2 U(q_c) \equiv -\beta^2 U(q_m)\equiv -E^{r}_{m}
\label{conserve}
\ee
Therefore, the instanton action is
\begin{eqnarray}
S[q_c] &=&
\int_{-L/2}^{+L/2} dt \left({1\over4} \dot
 q_{c}^{2}(t) +
{\beta^2\over 4} U(q_c) \right) \nonumber\\
&=& LE^{r}_{m}+
\int_{-L/2}^{+L/2} dt {1\over2} \dot q_{c}^{2}
\;\rightarrow\; LE^{r}_{m}+ \\
\lefteqn{+\int_{q_0}^{q_m} dq \sqrt{\beta^2 (U(q) - U(q_m))} \equiv LE^{r}_{m}+
 S_i}
\nonumber
\end{eqnarray}
The integration around the saddle point corresponding to $q_c$ in the gaussian
approximation provides the instanton contribution to the partition function.
This calculation involves a Jacobian factor, which becomes complex because of
the turning point $q_0$, and is proportional to $L$ because of traslation
invariance in the euclidean time. The final result is
\begin{eqnarray}
Im\ tr(e^{-LH}) &\propto& Le^{-S[q_c]}=Le^{-LE_{m}^{r} -S_i} \nonumber\\
&\propto& Im\  exp(-L(E_m^r + i e^{-S_i}))
\label {trazaccion}
\end{eqnarray}
Therefore, $e^{-S_i}$ corresponds to the imaginary part of the eigenvalue
whose real part is the value of the potential at $q_m$, $E=E_m^r + ie^{-S_i}$,
which corresponds to a metastable state

Let us particularize this formula to the case of pure gravity, eqs.(9,24),
where %
\begin{eqnarray}
&U(q) = \beta^{-{6\over5}} v(z) \quad ; \qquad v(z)=Az^3 -BTz & \nonumber\\
&q=\sqrt{4R_c} + \beta^{-{2\over5}} z&
\end{eqnarray}
and $E_m^r$ is just the semiclassical value of the Fermi energy.
Then, $S_i$ becomes finite in the double scaling  limit
 ($\beta\rightarrow\infty$)
and the result is
\be
S_i = {12\over 5} \root 4 \of {B^5 \over 27 A^3} T^{5\over4}
\label{accionA}
\ee
Therefore, the lifetime of this metastable state is
\be
\tau \propto e^{-S_i} \;\; ; \;\;\; S_i = {4\sqrt6\over5} X_c(\gamma
R_c^2)^{-{1\over4}} T^{5\over4}
\label{eq:lifetime}
\ee

As mentioned above, no related effect has been observed in the numerical
 computation
\cite{AmbG} . This negative result could be related to the numerical precission
 . In
\cite{AmbG} the the $D=0$ specific heat is  computed by introducing a source in
 the
Fokker--Planck potential,  $U_{FP}(\lambda) +J\lambda^2$. In this way, the
derivative of the corresponding free energy with respect to the source provides
$<Tr\ \Phi^2>$, which is related to the first derivative of the pure gravity
free energy, $<Tr\ \Phi^2>\propto F'=-<P>$. The consistency of this method
requires that the value of $J$ is small enough to keep the phase structure of
 pure
gravity. Implementing this for the generic even matrix model potential of
fourth
degree shows \cite{instabilities} that
in the double scaling limit, $\beta\rightarrow\infty$, this requires that the
source scales, $J=\beta^{-{4\over  5}} j$ and
if this scaling is not explicitly performed, the constraint on the source $J$
fixes the precission to be a function of $T$ decreasing substantially at small
 values of
T. In particular, for  the range explored in \cite{AmbG}, to preserve the
 precision
achieved at large $T$  when $T$ is small, one should increase the value of $N$
 with a
factor of  three. This is precisely the region where the instanton--like
 behaviour
should be observed.

Let us finally consider eq.\ref{eq:ufinal} for the next multicritical model,
$k=3$. Notice that in general, eq.\ref{eq:ufinal}
only gives the
leading terms of $u_{SC}$ in the double scaling limit, but it is exact when the
 {\it
canonical} representative is chosen for the potential. When $k=3$ this means a
potential of degree $6$. In such case, the result is
\begin{eqnarray}
u_{SC}(p)&\propto& \left(4R(X)-p^2\right)^{1\over2} \left(\left(p^2 + R(X) -
5R_c\right)^2 + \right. \nonumber\\
& & \left. + 5 \left(R(X)-R_c\right)^2\right)
\label{k3}
\end{eqnarray}
which means that, in this case, $U_{FP}(\lambda)$ has only one absolute
minimum at $\lambda=0$. Therefore, no metastable states are present in the
$k=3$ model, and no imaginary part is expected. This, of course,
agrees with the fact that the $k=3$ model is
already well defined as a matrix model.

We have been considering
non--perturbative effects  in the $ D=1-like$ spectrum of the doubled scaled
Fokker--Planck potential which stabilizes the matrix model, whose original
Borel--nonsummability is refleceted in the appearance of real instantons from
 the
secondary minima of the potential. The problem is how to translate these
results
 into
the actual $D=0$ problem. This  requires the computation of the
 non--perturbative
contributions to the $D=1$ density, not yet completed, which leaves
the interpretation of this instability open at present. On the other hand, one
 can
clearly stablish \cite{instabilities} that the tunneling scales properly,
 remaining
finite in the limit and that it involves the instanton action  in
eq. 33.

\vspace{.5 cm}
\noindent
4)CONCLUSIONS

We have analyzed the problem of the non--perturbative definition
of $2D$ pure gravity in $k$--even matrix models with the Fokker--Planck
stabilization, discussing other proposals. We have proved that the
fermionic formulation, on which previous results were based, is
only valid for $k<4$ , but that with a mean field in the WKB
approximation the generic  potential decouples and it is explicitly related
to the density of states of the matrix model.
We have shown then how to analyze analyticallly the critical behaviour of the
corresponding Fokker--Planck potential performing explicitly the double scaling
limit It turns out that the scaled potential has secondary minima in the
relevant $k$ even case, which exhibit instanton-like behaviour corresponding
to the Fermi level becoming complex and which could reflect the non--Borel
sumability of the pure gravity series solution of the Painlev\'e equation.
We have discussed how this result can be seen in one of the previous numerical
computations and why is not seen in the other.
The final interpretation of these instabilities in terms of sum over surfaces,
or the pure gravity true vacuum, requires more work as mentioned at the end
of last section, which is in progress.
\begin{acknowledge}
We thank J. Ambj{\o}rn, M. Asorey, J.Gonzalez, J. Jurkiewicz, O. Lechtenfeld
and
 M.
Vozmediano for discussions. L.M. was Fullbright scholar.
\end{acknowledge}

\end{document}